\documentclass[10pt,a4paper,onecolumn]{article}
\usepackage{marginnote}
\usepackage{graphicx}
\usepackage{xcolor}
\usepackage{authblk,etoolbox}
\usepackage{titlesec}
\usepackage{calc}
\usepackage{tikz}
\usepackage{hyperref}
\hypersetup{colorlinks,breaklinks,
            urlcolor=[rgb]{0.0, 0.5, 1.0},
            linkcolor=[rgb]{0.0, 0.5, 1.0}}
\usepackage{caption}
\usepackage{tcolorbox}
\usepackage{amssymb,amsmath}
\usepackage{ifxetex,ifluatex}
\usepackage{seqsplit}
\usepackage{fixltx2e} 
\usepackage[
  backend=biber,
]{biblatex}
\bibliography{paper.bib}

\usepackage[top=3.5cm, bottom=3cm, right=1.5cm, left=1.0cm,
            headheight=2.2cm, reversemp, includemp, marginparwidth=4.5cm]{geometry}

\titleformat{\section}
  {\normalfont\sffamily\Large\bfseries}
  {}{0pt}{}
\titleformat{\subsection}
  {\normalfont\sffamily\large\bfseries}
  {}{0pt}{}
\titleformat{\subsubsection}
  {\normalfont\sffamily\bfseries}
  {}{0pt}{}
\titleformat*{\paragraph}
  {\sffamily\normalsize}

\usepackage{fancyhdr}
\makeatletter
\let\ps@plain\ps@fancy
\fancyheadoffset[L]{4.5cm}
\fancyfootoffset[L]{4.5cm}

\definecolor{linky}{rgb}{0.0, 0.5, 1.0}

\newtcolorbox{repobox}
   {colback=red, colframe=red!75!black,
     boxrule=0.5pt, arc=2pt, left=6pt, right=6pt, top=3pt, bottom=3pt}

\newcommand{\ExternalLink}{%
   \tikz[x=1.2ex, y=1.2ex, baseline=-0.05ex]{%
       \begin{scope}[x=1ex, y=1ex]
           \clip (-0.1,-0.1)
               --++ (-0, 1.2)
               --++ (0.6, 0)
               --++ (0, -0.6)
               --++ (0.6, 0)
               --++ (0, -1);
           \path[draw,
               line width = 0.5,
               rounded corners=0.5]
               (0,0) rectangle (1,1);
       \end{scope}
       \path[draw, line width = 0.5] (0.5, 0.5)
           -- (1, 1);
       \path[draw, line width = 0.5] (0.6, 1)
           -- (1, 1) -- (1, 0.6);
       }
   }

\patchcmd{\@maketitle}{center}{flushleft}{}{}
\patchcmd{\@maketitle}{center}{flushleft}{}{}
\patchcmd{\@maketitle}{\LARGE}{\LARGE\sffamily}{}{}
\def\maketitle{{%
  
  \AB@maketitle}}
\makeatletter
\renewcommand\AB@affilsepx{ \protect\Affilfont}
\renewcommand\AB@affilnote[1]{{\bfseries #1}\hspace{3pt}}
\makeatother

\renewcommand\Affilfont{\sffamily\small\mdseries}
\ifnum 0\ifxetex 1\fi\ifluatex 1\fi=0 
  \usepackage[T1]{fontenc}
  \usepackage[utf8]{inputenc}

\else 
  \ifxetex
    \usepackage{mathspec}
  \else
    \usepackage{fontspec}
  \fi
  \defaultfontfeatures{Ligatures=TeX,Scale=MatchLowercase}

\fi
\IfFileExists{upquote.sty}{\usepackage{upquote}}{}
\IfFileExists{microtype.sty}{%
\usepackage{microtype}
\UseMicrotypeSet[protrusion]{basicmath} 
}{}

\usepackage{hyperref}
\hypersetup{unicode=true,
            pdftitle={OctApps: a library of Octave functions for continuous gravitational-wave data analysis},
            pdfborder={0 0 0},
            breaklinks=true}
\usepackage{graphicx,grffile}
\makeatletter
\def\maxwidth{\ifdim\Gin@nat@width>\linewidth\linewidth\else\Gin@nat@width\fi}
\def\maxheight{\ifdim\Gin@nat@height>\textheight\textheight\else\Gin@nat@height\fi}
\makeatother
\setkeys{Gin}{width=\maxwidth,height=\maxheight,keepaspectratio}
\IfFileExists{parskip.sty}{%
\usepackage{parskip}
}{
\setlength{\parindent}{0pt}
\setlength{\parskip}{6pt plus 2pt minus 1pt}
}
\providecommand{\tightlist}{%
  \setlength{\itemsep}{0pt}\setlength{\parskip}{0pt}}
\ifx\paragraph\undefined\else
\let\oldparagraph\paragraph
\renewcommand{\paragraph}[1]{\oldparagraph{#1}\mbox{}}
\fi
\ifx\subparagraph\undefined\else
\let\oldsubparagraph\subparagraph
\renewcommand{\subparagraph}[1]{\oldsubparagraph{#1}\mbox{}}
\fi

\title{OctApps: a library of Octave functions for continuous gravitational-wave
data analysis}

        \author[1, 2, 3]{Karl Wette}
          \author[2, 3]{Reinhard Prix}
          \author[4, 2, 3]{David Keitel}
          \author[4]{Matthew Pitkin}
          \author[2, 3]{Christoph Dreissigacker}
          \author[5, 6]{John T. Whelan}
          \author[7, 8, 6]{Paola Leaci}
    
      \affil[1]{ARC Centre of Excellence for Gravitational Wave Discovery (OzGrav) and
Centre for Gravitational Physics, Research School of Physics and
Engineering, The Australian National University, ACT 0200, Australia}
      \affil[2]{Max-Planck-Institut für Gravitationsphysik (Albert-Einstein-Institut),
D-30167 Hannover, Germany}
      \affil[3]{Leibniz Universität Hannover, D-30167 Hannover, Germany}
      \affil[4]{Institute for Gravitational Research, SUPA, University of Glasgow,
Glasgow G12 8QQ, U.K.}
      \affil[5]{School of Mathematical Sciences and Center for Computational Relativity
\& Gravitation, Rochester Institute of Technology, Rochester, NY 14623,
U.S.A.}
      \affil[6]{Max-Planck-Institut für Gravitationsphysik (Albert-Einstein-Institut),
D-14476 Golm, Germany}
      \affil[7]{Università di Roma `La Sapienza', I-00185 Roma, Italy}
      \affil[8]{INFN, Sezione di Roma, I-00185 Roma, Italy}
  \date{\vspace{-5ex}}

\begin{document}
\maketitle

\marginpar{
  \sffamily\small

  {\bfseries DOI:} \href{https://doi.org/10.21105/joss.00707}{\color{linky}{10.21105/joss.00707}}

  \vspace{2mm}

  {\bfseries Software}
  \begin{itemize}
    \setlength\itemsep{0em}
    \item \href{https://github.com/openjournals/joss-reviews/issues/707}{\color{linky}{Review}} \ExternalLink
    \item \href{https://github.com/octapps/octapps}{\color{linky}{Repository}} \ExternalLink
    \item \href{http://dx.doi.org/10.5281/zenodo.1283525}{\color{linky}{Archive}} \ExternalLink
  \end{itemize}

  \vspace{2mm}

  {\bfseries Submitted:} 10 April 2018\\
  {\bfseries Published:} 06 June 2018

  \vspace{2mm}
  {\bfseries Licence}\\
  Authors of papers retain copyright and release the work under a Creative Commons Attribution 4.0 International License (\href{http://creativecommons.org/licenses/by/4.0/}{\color{linky}{CC-BY}}).
}

\section{Summary}\label{summary}

Gravitational waves are minute ripples in spacetime, first predicted by
Einstein's general theory of relativity in 1916. Their existence has now
been confirmed by the recent successful detections of gravitational
waves from the collision and merger of binary black holes (Abbott 2016)
and binary neutron stars (Abbott 2017) in data from the
\href{https://www.ligo.org}{LIGO} and
\href{http://www.virgo-gw.eu}{Virgo} gravitational-wave detectors.
Gravitational waves from rapidly-rotating neutron stars, whose shape
deviates from perfect axisymmetry, are another potential astrophysical
source of gravitational waves, but which so far have not been detected.
The search for this type of signals, also known as continuous waves,
presents a significant data analysis challenge, as their weak signatures
are expected to be buried deep within the instrumental noise of the LIGO
and Virgo detectors. For reviews of continuous-wave sources, data
analysis techniques, and recent searches of LIGO and Virgo data, see for
example Prix (2009) and Riles (2017).

The \emph{OctApps} library provides various functions, written in Octave
(Eaton et al. 2015), intended to aid research scientists who perform
searches for continuous gravitational waves. They are organized into the
following directories:

\begin{itemize}
\tightlist
\item
  \texttt{src/cw-data-analysis}: general-purpose functions for
  continuous-wave data analysis.
\item
  \texttt{src/cw-line-veto}: functions which implement detection
  statistics which are robust to instrumental disturbances in the
  detector data, as described in (Keitel et al. 2014).
\item
  \texttt{src/cw-metric-template-banks}: functions which determine the
  number of filtering operations required to search for continuous waves
  over various astrophysical parameter spaces, described further in
  (Wette and Prix 2013) and (Leaci and Prix 2015).
\item
  \texttt{src/cw-optimal-search-setup}: functions which determine the
  optimally-sensitive search for continuous gravitational waves, given a
  fixed computing budget, following the method of (Prix and Shaltev
  2012).
\item
  \texttt{src/cw-sensitivity}: functions which predict the sensitivity
  of a search for continuous waves, following the method of (Wette
  2012).
\item
  \texttt{src/cw-weave-models}: functions which characterize the
  behaviour of \emph{Weave}, an implementation of an optimized search
  pipeline for continuous waves (Wette et al. 2018).
\end{itemize}

Many of these scripts make use of C functions from the
\href{https://wiki.ligo.org/DASWG/LALSuite}{LSC Algorithm Library Suite
(LALSuite)}, using \href{http://www.swig.org}{SWIG} to provide the
C-to-Octave interface.

In addition, \emph{OctApps} provides various general-purpose functions,
which may be of broader interest to users of Octave, organized into the
following directories:

\begin{itemize}
\tightlist
\item
  \texttt{src/array-handling}: manipulation of Octave arrays and cell
  arrays.
\item
  \texttt{src/command-line}: includes \texttt{parseOptions()}, a
  powerful parser for Octave function argument lists in the form of
  key--value pairs. Together with \texttt{octapps\_run}, a Unix shell
  script, it allows Octave functions to also be called directly from the
  Unix command line using a \texttt{-\/-key=value} argument syntax.
\item
  \texttt{src/condor-jobs}: submission of jobs to a computer cluster
  using the \href{https://research.cs.wisc.edu/htcondor}{HTCondor} job
  submission system. It includes \texttt{depends()}, a low-level
  function written using Octave's C++ API which, given the name of an
  Octave function, returns the names of all Octave functions called by
  the named function; it is used to deploy a self-contained tarball of
  Octave \texttt{.m} files to a remote node on a computer cluster.
\item
  \texttt{src/convert-units}: functions which convert between different
  angular units, and between different time standards.
\item
  \texttt{src/file-handling}: parsing of various file formats, such as
  FITS and \texttt{.ini}.
\item
  \texttt{src/general}: miscellaneous general-purpose functions.
\item
  \texttt{src/geometry}: mathematical operations associated with
  geometric objects, e.g.~computing the intersection of two lines.
\item
  \texttt{src/histograms}: includes \texttt{@Hist}, an Octave class
  representing a histogram, with various method functions which perform
  common statistical operations, e.g.~computing the cumulative
  distribution function.
\item
  \texttt{src/lattices}: mathematical operations associated with lattice
  theory, e.g.~computing the nearest point in a lattice to a given point
  in space.
\item
  \texttt{src/mathematical}: miscellaneous general mathematical
  functions, including some C functions incorporated from the GNU
  Scientific Library (Galassi 2009), using SWIG to provide the
  C-to-Octave interface.
\item
  \texttt{src/plotting}: helper functions for plot creation and output
  in \href{https://www.tug.org}{TeX} format.
\item
  \texttt{src/statistics}: miscellaneous statistical functions,
  particularly for probability distributions.
\item
  \texttt{src/text-handling}: various functions for creating formatted
  text output.
\item
  \texttt{src/version-handling}: handling of version information,
  particularly from the \href{https://git-scm.com}{Git} version control
  system.
\end{itemize}

Development of \emph{OctApps} is hosted on
\href{https://github.com/octapps/octapps}{GitHub}; a test suite of all
functions in \emph{OctApps} is regularly integrated on
\href{https://travis-ci.org/octapps/octapps}{Travis CI}. The
\href{https://github.com/octapps/octapps/blob/master/README.md}{README}
file provides instructions for building, testing, and contributing to
\emph{OctApps}, as well as a full list of prerequisite software required
by \emph{OctApps}. A \href{https://octapps.github.io}{reference manual}
for \emph{OctApps} in HTML format is available; documentation of each
\emph{OctApps} function can also be accessed through the \texttt{help}
function in Octave.

\subsection{Acknowledgements}\label{acknowledgements}

We acknowledge that \emph{OctApps} includes contributions from Zdravko
Botev, Ronaldas Macas, John McNabb, and Daniel de Torre Herrera. We
thank Steven R. Brandt for reviewing this paper and providing the
\texttt{Dockerfile} for building and testing \emph{OctApps}. This work
was supported by the Max Planck Society, by German Research Foundation
(DFG) grant SFB/TR 7, and by the German Aerospace Center (DLR). KW is
supported by Australian Research Council (ARC) grant CE170100004. MP is
funded by the UK Science \& Technology Facilities Council (STFC) under
grant ST/N005422/1. This paper has document number LIGO-P1800078-v4.

\section*{References}\label{references}
\addcontentsline{toc}{section}{References}

\hypertarget{refs}{}
\hypertarget{ref-LIGOVirg2016a}{}
Abbott, B. P. et al. 2016. ``Observation of Gravitational Waves from a
Binary Black Hole Merger.'' \emph{Physical Review Letters} 116: 061102.
doi:\href{https://doi.org/10.1103/PhysRevLett.116.061102}{10.1103/PhysRevLett.116.061102}.

\hypertarget{ref-LIGOVirg2017a}{}
---------. 2017. ``GW170817: Observation of Gravitational Waves from a
Binary Neutron Star Inspiral.'' \emph{Physical Review Letters} 119:
161101.
doi:\href{https://doi.org/10.1103/PhysRevLett.119.161101}{10.1103/PhysRevLett.119.161101}.

\hypertarget{ref-Octave2015}{}
Eaton, John W., David Bateman, Søren Hauberg, and Rik Wehbring. 2015.
\emph{GNU Octave Version 4.0.0 Manual: A High-Level Interactive Language
for Numerical Computations}.
\url{http://www.gnu.org/software/octave/doc/interpreter}.

\hypertarget{ref-GSL2009}{}
Galassi, M. et al. 2009. \emph{GNU Scientific Library Reference Manual -
Third Edition}. \url{https://www.gnu.org/software/gsl}.

\hypertarget{ref-KeitEtAl2014a}{}
Keitel, D., R. Prix, M. A. Papa, P. Leaci, and M. Siddiqi. 2014.
``Search for continuous gravitational waves: Improving robustness versus
instrumental artifacts.'' \emph{Physical Review D} 89: 064023.
doi:\href{https://doi.org/10.1103/PhysRevD.89.064023}{10.1103/PhysRevD.89.064023}.

\hypertarget{ref-LeacPrix2015a}{}
Leaci, P., and R. Prix. 2015. ``Directed searches for continuous
gravitational waves from binary systems: Parameter-space metrics and
optimal Scorpius X-1 sensitivity.'' \emph{Physical Review D} 91: 102003.
doi:\href{https://doi.org/10.1103/PhysRevD.91.102003}{10.1103/PhysRevD.91.102003}.

\hypertarget{ref-Prix2009a}{}
Prix, R. 2009. ``Gravitational Waves from Spinning Neutron Stars.'' In
\emph{Neutron Stars and Pulsars}, edited by W. Becker, 357:651.
Astrophysics and Space Science Library. Berlin/Heidelberg: Springer.
doi:\href{https://doi.org/10.1007/978-3-540-76965-1_24}{10.1007/978-3-540-76965-1\_24}.

\hypertarget{ref-PrixShal2012a}{}
Prix, R., and M. Shaltev. 2012. ``Search for continuous gravitational
waves: Optimal StackSlide method at fixed computing cost.''
\emph{Physical Review D} 85: 084010.
doi:\href{https://doi.org/10.1103/PhysRevD.85.084010}{10.1103/PhysRevD.85.084010}.

\hypertarget{ref-Rile2017a}{}
Riles, K. 2017. ``Recent searches for continuous gravitational waves.''
\emph{Modern Physics Letters A} 32: 1730035--1730685.
doi:\href{https://doi.org/10.1142/S021773231730035X}{10.1142/S021773231730035X}.

\hypertarget{ref-Wett2012a}{}
Wette, K. 2012. ``Estimating the sensitivity of wide-parameter-space
searches for gravitational-wave pulsars.'' \emph{Physical Review D} 85:
042003.
doi:\href{https://doi.org/10.1103/PhysRevD.85.042003}{10.1103/PhysRevD.85.042003}.

\hypertarget{ref-WettPrix2013a}{}
Wette, K., and R. Prix. 2013. ``Flat parameter-space metric for all-sky
searches for gravitational-wave pulsars.'' \emph{Physical Review D} 88:
123005.
doi:\href{https://doi.org/10.1103/PhysRevD.88.123005}{10.1103/PhysRevD.88.123005}.

\hypertarget{ref-Wett2018a}{}
Wette, K., S. Walsh, R. Prix, and M. A. Papa. 2018. ``Weave: a
semicoherent search implementation for continuous gravitational waves.''
\emph{Submitted to Physical Review D}.

\end{document}